\documentclass[twocolumn,groupedaddress,superscriptaddress,showpacs,prb,letterpaper]{revtex4}

\usepackage{times,amsmath,amssymb}	
\usepackage{graphicx} % Include figure files
\usepackage{bm} % bold math

\begin{document}

%************************************************************
% Begining of the Manuscript
%************************************************************

%-----------------------------------------------------------
% Front matter
%-----------------------------------------------------------

\title{Conductance quantization in mesoscopic graphene}


\date{\today}

\author{N.~M.~R. Peres}
\affiliation{Center of Physics and Departamento de F{\'\i}sica,
Universidade do Minho, P-4710-057, Braga, Portugal}

\author{A.~H. Castro Neto}
\affiliation{Department of Physics, Boston University, 590 
Commonwealth Avenue, Boston, MA 02215, USA}

\author{F. Guinea}
\affiliation{Instituto de  Ciencia de Materiales de Madrid, CSIC,
 Cantoblanco E28049 Madrid, Spain}

%-----------------------------------------------------------
% Abstract
%-----------------------------------------------------------

\begin{abstract}
Using a generalized Landauer approach we study the non-linear
transport in mesoscopic graphene with zig-zag and armchair
edges. We find that for clean systems, the low-bias low-temperature
conductance, $G$, of an 
armchair edge system in quantized as $G/\tilde t=4 n e^2/h$,
whereas for a zig-zag edge the quantization changes
to  $G/\tilde t=4(n+1/2)e^2/h$, where $\tilde t$ is the transmission
probability and $n$ is an integer. We also study the effects of a 
non-zero bias, temperature, and magnetic field on the conductance. 
The magnetic field dependence of the quantization plateaus in 
these systems is somewhat different from the one found in the
two-dimensional electron gas due to a different Landau level quantization.  
\end{abstract}

\maketitle

%-----------------------------------------------------------
% Section - Introduction
%-----------------------------------------------------------

\section{Introduction}
\label{intro}

Graphene, a two-dimensional (2D) carbon system on a honeycomb
lattice presents many anomalous properties when compared with
the well-known 2D electron gas obtained in heterostructures. One
of the most striking properties is an exotic integer quantum Hall effect 
(IQHE)  
predicted theoretically \cite{pga,sg}, and 
measured recently  \cite{geim,kim}. The IQHE shows a Hall 
conductivity given by: $\sigma_{xy}=2(2n+1)e^2/h$, where $n$ is a 
positive integer. Interestingly, the electrical properties of the 
graphene systems can be considered classical, 
in the sense that the measured conductance of the systems is found to 
increase with the increase of system width and to decrease 
with the increase of the system length \cite{berger}. 
This experimental result
can be understood as an evidence for the presence of disorder in the 
measured systems. This is further supported 
by the difficulty in finding experimental evidence for
a fractional quantum Hall effect (FQHE) 
\cite{pga,pga_hall}.

\begin{figure}[hf]
\begin{center}
\includegraphics*[width=.99\columnwidth]{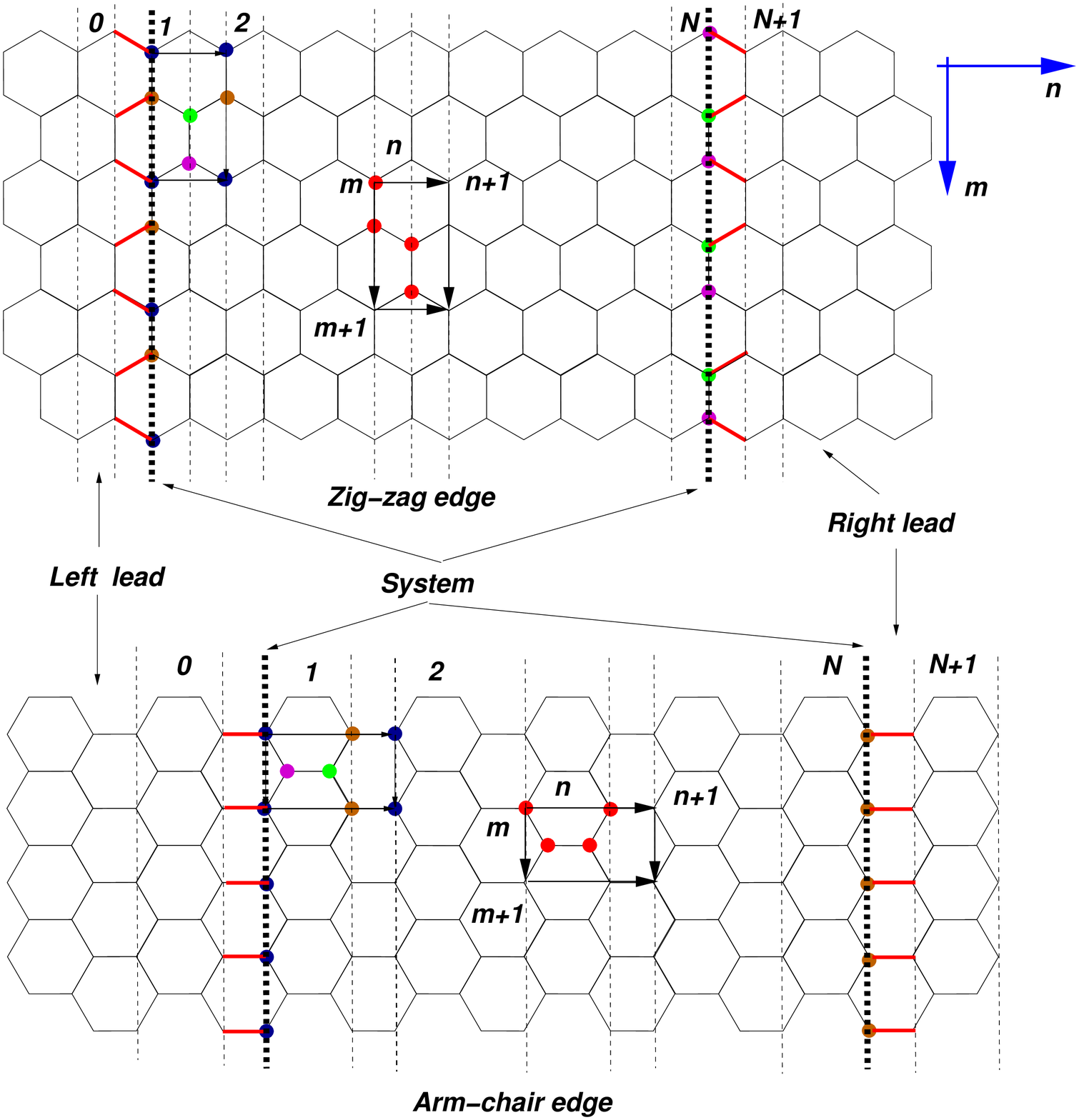}
\end{center}
\caption{(color online) Geometry of a finite-size honeycomb 
lattice. Top: Sample with a zig-zag edge; 
Bottom: Sample with an armchair edge. The thick dotted lines
represent the position of the leads which are assumed to
be made out of graphene. The rectangles in the top left corner
of the figures (close to the left lead) show the geometry of the unit cell.}
\label{transport_scheme}
\end{figure}

In the closely related field of carbon nanotubes,
recent experiments showed that the conductance of a single wall
carbon nanotube is quantized \cite{liang} and shows Fabry-Perot
interference patterns. These results can be explained within a 
generalized Landauer approach, $S-$matrix theory \cite{liang,dattabook}, 
and non-equilibrium Green's function methods \cite{keldysh,barnas}. 
The formulation of the problem was introduced 
by  Lake {\it et al.}\cite{lake}, after the work of Caroli {\it et al.}   
\cite{caroli,combescot,combescot&schreder}. Because carbon
nanotubes are essentially wrapped graphene, we expect
conductance quantization and Fabry-Perot
interference patterns to be also observable in ultra-clean graphene. 
The quantization and the interference patterns,
however, should reflect the different types of edges a graphene
sheet has (see Fig.~\ref{transport_scheme}).

The importance of zig-zag and armchair edges in graphene sheets
has been recognized in electron microscopy \cite{niimi}. 
Clearly, these two types of edges produce very different 
electron microscopy intensity curves. We expect, therefore, 
that coherent charge transport should
be different if measured in systems with different
types of edges. As in carbon nanotubes \cite{datta},
a simple Landauer approach to determine the quantization of 
the conductance, $G$, can be used for mesoscopic graphene sheets. 
The calculations follow
 the  generalization of the Landauer approach introduced earlier 
by Bagwell and Orlando \cite{orlando}. This type of approach
does not account for a discussion of interference patterns,
since it neglects multiple electronic reflection at the
contacts \cite{liang,barnas}, and it will be discussed elsewhere \cite{pga2}.

% ------------------------------------------------------------
% FIGURE 1 BEGINS
% ------------------------------------------------------------
% ------------------------------------------------------------

This paper is organized as follows: in Sec.~\ref{landauer}
we introduce the tight-binding solution for electrons in graphene
strips having two different types of edges and the differences 
and similarities between graphene and carbon
nanotubes are discussed. Starting from the tight binding solution of laterally confined
electron in graphene strips we introduce a Landauer analysis allowing for
the calculation of the conductance due to quasi-one-dimensional transport
arising from the confinement;  this is done at
at zero-temperature and in the zero-bias limit.
In Sec.~\ref{finiteB} we discuss the effect of temperature, gate bias, and 
magnetic field  on the conductance curves. It is found that
conductance has a minimum value as function of the gate potential
(or Fermi energy), from which a ``V-like'' curve
develops, in agreement with measured transport
properties in graphene samples. 
Sec.~\ref{sum} contains our conclusions.

%-----------------------------------------------------------
% Section - Introduction
%-----------------------------------------------------------

\section{A Landauer analysis}
\label{landauer}

The geometry of the problem is shown in Fig.~\ref{transport_scheme}, where the left and right  leads 
are supposed to be made of graphene \cite{barnas}. 
This aspect is not essential
in the approach we develop below, 
where the contacts to the leads will be modeled
by a transmission probability (as done in Refs.~[\onlinecite{liang,barnas}] in the context
of carbon nanotubes).
The systems studied here are assumed to have a very
asymmetric aspect ratio, where the length, $L$, is much 
larger than their width, $W$. These systems have some similarity
with carbon nanotubes \cite{datta} 
but differ from them in a fundamental way:
the absence of periodic boundary conditions along the direction
perpendicular to the edges (the $m$ direction). 
As a consequence, it is possible to have different kinds
of strips, characterized by different types of edges.
In what follows, we discuss the cases of zig-zag
and armchair edges although other edge geometries can be
studied with the same methods.

The calculation of the conductance $G$ following a Landauer type
of approach\cite{datta,orlando} requires the solution of a
tight-binding problem in a finite geometry. The tight-binding Hamiltonian has the form :
\begin{eqnarray}
H_{{\rm t.b.}} &=& -t \sum_{\langle i,j \rangle ,\sigma} (a^{\dag}_{i,\sigma} b_{j,\sigma}
+ {\rm h.c.}) \nonumber \\
&+& t' \sum_{\langle \langle i,j \rangle \rangle ,\sigma} (a^{\dag}_{i,\sigma} a_{j,\sigma}
+ b^{\dag}_{i,\sigma} b_{j,\sigma} + {\rm h.c.}) \, ,
\label{Htb}
\end{eqnarray} 
where $a^{\dag}_{i,\sigma}$ ($a_{i,\sigma}$) creates (annihilates) an
electron on site ${\bf R}_i$, with spin $\sigma$ 
($\sigma = \uparrow,\downarrow$) on sub-lattice $A$ and $b^{\dag}_{i,\sigma}$
($b_{i,\sigma}$)  creates (annihilates) and
electron on site ${\bf R}_i$ with spin $\sigma$ 
($\sigma = \uparrow,\downarrow$) on sub-lattice $B$. $t$ is the nearest
neighbor ($\langle i, j \rangle$) 
hopping energy ($t \approx 2.7$ eV), and $t'$ is the
next-nearest neighbor ($\langle \langle i,j \rangle \rangle$) 
hopping energy ($t'/t \approx 0.1$). In what follows we 
suppress the spin index since it plays no role 
(apart from a degeneracy factor). 

In an infinite system the Hamiltonian (\ref{Htb}) can be easily diagonalized and one
can show that the low energy electronic excitations of the problem
reside around the K-points of the Brillouin zone \cite{wallace}
and have a dispersion given by ($t'$ does not remove the
Dirac point):
\begin{eqnarray}
E_{\pm}({\bf k}) = \pm v_F |{\bf k}| \, ,
\label{dirac}
\end{eqnarray}
where ${\bf k} = (k_x,k_y)$ is a two-dimensional momentum,
and $v_F = 3 t a/(2\hbar)$ (where $a$ is the lattice spacing) is the
Fermi-Dirac velocity. Eq. (\ref{dirac}) describes the dispersion relation
of Dirac electrons with a speed $v_F$. One of the consequences
of the Dirac dispersion is that the fermions in the system have
zero effective mass, and a linearly vanishing 
density of states, $N(E)$ ($N(E) \propto |E|$), at low energies. The linearly 
dispersing electrons and the vanishing of
the density of states lead to a very anomalous metallic behavior
with many non-Fermi liquid properties \cite{pga}. These anomalous
properties are reflected in the experimentally measurable quantities, 
such as the Hall conductivity in the IQHE \cite{geim,kim}. We are
going to show that the presence of Dirac fermions in the spectrum
has also a strong effect in the conductance of finite graphene strips.

% ------------------------------------------------------------
% FIGURE 2 BEGINS
% ------------------------------------------------------------
\begin{figure}[hf]
\begin{center}
\includegraphics*[width=.99\columnwidth]{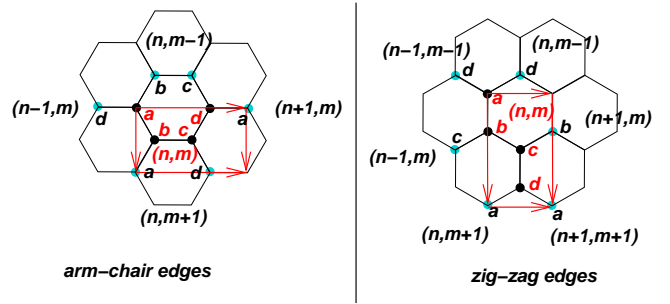}
\end{center}
\caption{(color online)
Label of the carbon atoms for Hamiltonian (\ref{Htb}) 
in the bulk of the system
(only the case $t'=0$ is shown for simplicity). The reader is referred to Fig.~\ref{transport_scheme}
where the unit cell is shown in each case. The black circles, labeled $a$, $b$, $c$, and $d$, inside the rectangle or
on the rectangle edges, belong
to same unit cell; the other circles represent lattice points
in adjacent unit cells connect by the hopping matrix $t$. }
\label{TB_bulk}
\end{figure}
% ------------------------------------------------------------

% ------------------------------------------------------------
% FIGURE 3 BEGINS
% ------------------------------------------------------------
\begin{figure}[hf]
\begin{center}
\includegraphics*[width=.99\columnwidth]{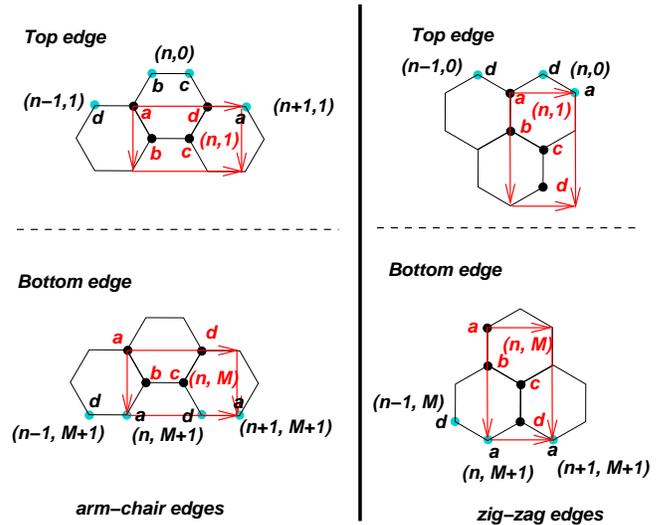}
\end{center}
\caption{(color online)
This figure shows how to deal with the tight-binding 
boundary conditions introduced by the free edges of the sample.
The cells on the right refer to top free edge and the bottom free
edge of a zig-zag sample (top of Fig.~\ref{transport_scheme} ).
The cells on the left refer to the top free edge and the bottom free
edge of an arm-chair sample (bottom of Fig.~\ref{transport_scheme} ).
As in Fig.~\ref{TB_bulk}, only the sites connected by $t$ are shown.}
\label{TB_edges}
\end{figure}
% ------------------------------------------------------------

In a finite system the boundary conditions (and hence, the
type of edges) become important in order to define the energy spectrum. In Fig.~\ref{TB_bulk}
we establish the labeling of the carbon sites in the bulk of the 
honeycomb lattice, and in Fig.~\ref{TB_edges} we present the 
labels for the carbon atoms close to the free edges of the system.
Notice that each unit cell in Fig.~\ref{TB_bulk} and Fig.~\ref{TB_edges} 
contains four carbon atoms,  
labeled by wave function amplitudes
$a(n,m)$, $b(n,m)$, $c(n,m)$, and $d(n,m)$, where $n$ and $m$
are integers that label each unit cell.
Since we are assuming that in each system only the edges
are different, we impose periodic boundary conditions along the
direction parallel to the edges (the $n$ direction,) 
leading to one-dimensional (1D) transport where the electronic states can be labeled
by the momentum $q_x$. If $N_y$ is the number of unit cells 
along the $m$ direction, then the
the tight-binding problem with zig-zag edges has 
a dimension given by
$(4N_y+2)\times (4N_y+2)$, while for the armchair
edge its dimension is
$(4N_y+4)\times (4N_y+4)$. 
For a nanotube, with periodic boundary
conditions, the problem can be formulated in 
terms of two amplitudes instead of four.

\begin{figure}[hf]
\begin{center}
\includegraphics*[width=.99\columnwidth]{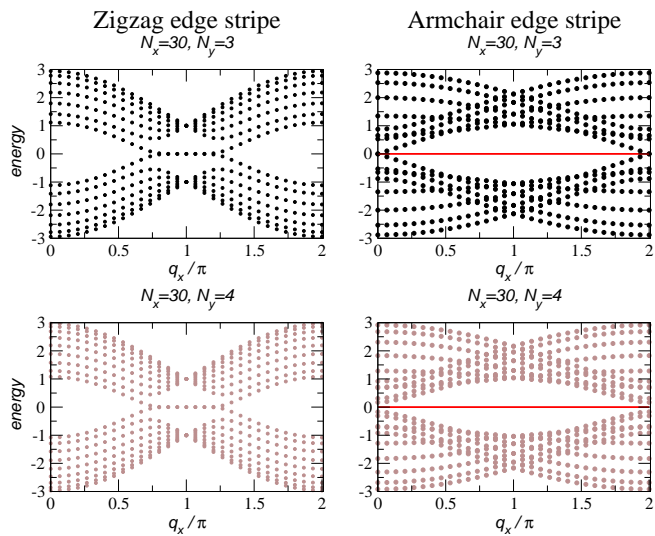}
\end{center}
\caption{(color online)
Energy levels for $t'=0$ (in units of $t$) as 
function of $q_x$ (in units of the lattice spacing
along the $n$ direction) for the two 
different types of edges. $N_y$ is the number of unit cells 
 perpendicular to the edges
(along the $m$ direction) and  $N_x$ gives the number of
momenta used in the abscissa (it also corresponds to the number
of unit cells along the $n$ direction).
The horizontal line at zero energy on the right panels has
an interception with zero energy modes in the upper panel
(metallic behavior) and no interception in the
lower panel (semi-conducting behavior).}
\label{TB_energy}
\end{figure}

The calculation of the conductance of a two-dimensional lattice
system with a very asymmetric aspect ratio \cite{datta} requires the 
identification of the number of transverse modes, $M(\epsilon)$, at a given energy,
$\epsilon$. $M(\epsilon)$ can be obtained from the solution of  the corresponding 
tight-binding problem. In Fig.~\ref{TB_energy} we show the energy
bands obtained from the diagonalization of the tight-binding
Hamiltonian (\ref{Htb}) with $t'=0$ for the zig-zag and armchair edges. 
It is clear that the two different types of edges lead to two 
very different band structures, especially close to zero energy.

% ------------------------------------------------------------
% FIGURE 4 BEGINS
% ------------------------------------------------------------
% ------------------------------------------------------------

It is important to mention the similarities and the differences
between  planar systems \cite{dresselhaus_edges} and 
carbon nanotubes \cite{dresselhaus_tubules}.
An armchair nanotube 
has the hexagons having two sides perpendicular to the
tube axis, in the zig-zag nanotube the hexagons have two sides
parallel to the tube axis. A graphene sample with a zig-zag 
edge has an energy spectrum presenting some
similarities with an armchair nanotube. It has two
bands crossing the chemical potential at zero energy
(and finite momentum).
A graphene sample with an armchair edge has a gap
at zero energy (near $q_x=0$) and hence is insulating, except when
$N_y$, the number of unit cells in the $m$-direction, is a multiple of
3, in which case the gap goes to zero and the material is
metallic (in fact a zero gap semi-conductor;
see Fig.~\ref{TB_energy})\cite{dresselhaus_edges}.

A generalized Landauer
approach \cite{orlando} shows that the tunneling current
is given by: 
\begin{equation}
I(V,T)=\frac {2e}h\int d\epsilon M(\epsilon)
\tilde t(\epsilon,V)[f(\epsilon-\mu_1)-
f(\epsilon-\mu_2)]\,,
\label{I}
\end{equation}
where  
$\tilde t(\epsilon,V)$ is the transmission probability 
per conducting mode at the energy 
$\epsilon$, $f(\epsilon) = 1/(e^{\epsilon/T}+1)$ is the 
Fermi-Dirac distribution ($T$ is the temperature and we 
have put
$k_B=1$),
 $V$ is the bias voltage applied to the system, and $\mu_1$ ($\mu_2$) 
is the chemical potential at right (left) lead 
($\mu_1=\mu_2 + eV$).

In a clean system, all 1D modes can carry electric current,
as long as they have a finite velocity in the direction parallel
to the edge. 
Although the zig-zag edge system has zero
energy modes with finite $q_x$ momentum, the group velocity
of these modes is zero when $t'=0$, and therefore they do not
contribute to the conductance. 
If we neglect the effect of the next
nearest neighbor hopping ($t'=0$),
both edge systems have two conducting zero energy modes
(choosing a metallic armchair edge system--$N_y$ multiple of three), and as a
consequence the small bias conductance is given by 
$4e^2 \tilde t/h$, at zero chemical potential. 

\begin{figure}[hf]
\begin{center}
\includegraphics*[width=.99\columnwidth]{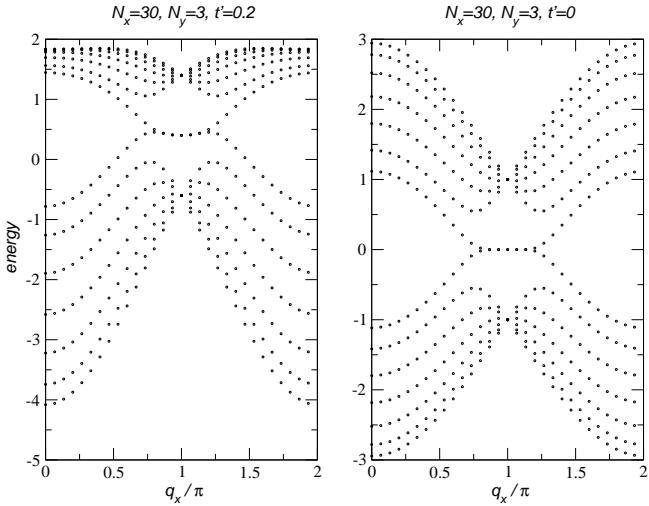}
\end{center}
\caption{(color online)
Energy levels (in units of $t$) as function of $q_x$ 
(in units of the lattice spacing
along the $n$ (or $x$) direction)
for a zig-zag system. Right: $t'=0$; left: $t' = 0.2 t$ }
\label{TB_tp}
\end{figure}

With the addition of a next nearest neighbor hopping ($t' \neq 0$)  
the picture
for the zig-zag edge changes substantially: (1) the half-filling case 
occurs at finite energy (not at zero energy); (2) the
flat band states located at zero energy, for $t'=0$, become
dispersive for $t'\ne 0$, and therefore the conductance is
modified. Fig.~\ref{TB_tp} shows the energy levels as function
of the momentum $q_x$ for a zig-zag system with $t'=0$
and $t' = 0.2 t$, for $N_y=3$ (this rather small $N_y$ value
allows the individual visualization of the transverse modes over the full
bandwidth, as it does in Fig.~\ref{TB_energy} as well). 

% ------------------------------------------------------------
% FIGURE 5 BEGINS
% ------------------------------------------------------------
% ------------------------------------------------------------

\begin{figure}[ht]
\begin{center}
\includegraphics*[width=.99\columnwidth]{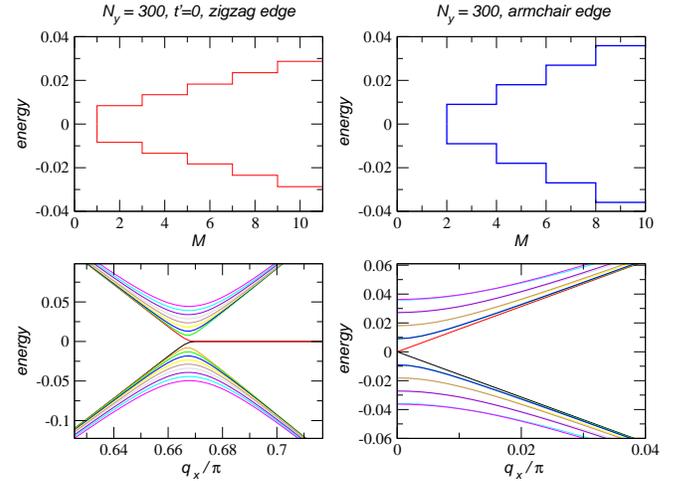}
\end{center}
\caption{(color online)
Lower panels: one-dimensional energy bands 
(energy in units of $t$) associated with a zig-zag (left) 
and a armchair edge (right) systems for $t'= 0$. 
Upper panels: number of 1D channels, $M$, as a
function of energy (in units of $t$).}
\label{TB_M1}
\end{figure}

In order to determine the conductance it is necessary to know
how  many 1D transverse modes are active for 
transport at a given energy.  For an armchair edge with 
 $N_y$ multiple of 3 (and hence a metallic system) one 
has two conducting modes at zero energy (see Fig.~\ref{TB_M1}) . 
As one moves away from zero energy the number of modes increases 
 as shown in Fig.~\ref{TB_M1}.  Hence, the
zero-bias zero-temperature conductance of a clean metallic armchair
system is given by: 
\begin{equation}
G=\frac {4ne^2}{h}\tilde t\,,
\label{Garm}
\end{equation}
with $n$ a positive integer. The value of $n$ depends on the 
value of the gate
potential controlling the  electronic density. The conductance also depends on 
$\tilde t$, the transmission probability, assumed to be energy
independent for simplicity.

A zig-zag clean system with $t'=0$ shows a different dependence.  
At zero energy $G$ is given by  $4e^2\tilde t/h$ because of the presence
of the two conducting zero-energy modes (see Fig.~\ref{TB_M1}).
However, as the gate potential
moves slightly away from zero the value of the conductance drops 
to $2e^2\tilde t/h$, since only one transverse mode is available.  
So the situation with a zig-zag system is somewhat ill-defined.
It reasonable to expect that this sudden change on the conductance
is difficult to be experimentally observed, since it would require an extreme fine tuning of the
experimental parameters. Therefore, we expect that  
the zero-bias zero-temperature conductance of a system with
a zig-zag edge to be given by: 
\begin{equation}
G=\frac {2(2n+1)e^2}{h}\tilde t\,,
\label{Gzig}
\end{equation}
with $n$ depending also on the value of the gate
potential. 

\begin{figure}[ht]
\begin{center}
\includegraphics*[width=.99\columnwidth]{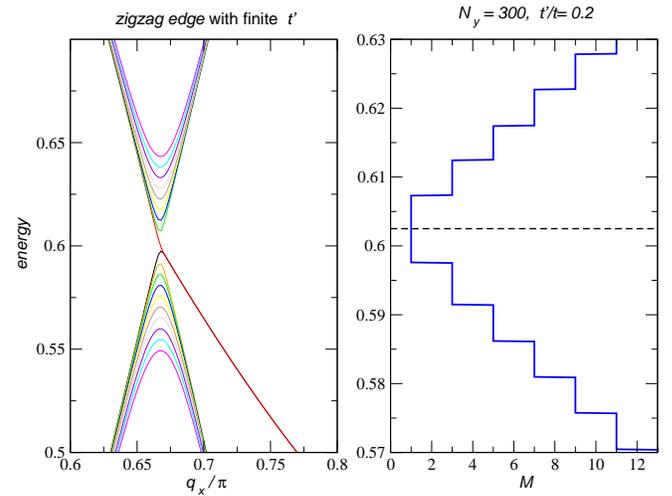}
\end{center}
\caption{(color online)
Left panel: one-dimensional energy band 
(energy in units of $t$), for a zig-zag edge system with $t' = 0.2 t$. 
Right panel: number of 1D channels, $M$, as a function of energy
(in units of $t$). The horizontal dashed line helps in stressing the
fact that $M(\epsilon)$ is asymmetric as a consequence of a finite $t'$.}
\label{TB_M2}
\end{figure}

The special value of $G$ we found  exactly at zero energy
(in the zig-zag case) 
does not survive, however, when
electron-hole symmetry (by this it is understood that the energy spectrum
is not symmetric around zero energy)
is broken by the presence of a
finite $t'$. When $t' \neq 0$ the zero mode acquires dispersion
and the conductance is given by Eq. (\ref{Gzig}).
In Fig.~\ref{TB_M2}, the effect of $t'$ in the low-energy band structure and in $M(\epsilon)$ is visible. Clearly the zero energy modes
have been removed, leading to two degenerate dispersive bands,
and an asymmetry in energy in the steps of $M$ 
is introduced. The asymmetry in energy
of $M(\epsilon)$  is an experimental
way of measuring the value of $t'$ in these systems. In the 
armchair case, the effect of $t'$ on $G$
is not as dramatic as in the case of the zig-zag edge system,
since in this case the conductance does not have an abrupt change
around zero energy.
 
In a system with several graphene planes 
each plane contributes to the conductance almost independently
because of the weak coupling between graphene sheets. Therefore,
the resulting conductance should be, at least approximately,
given by the above results multiplied by the number of layers.

% ------------------------------------------------------------
% FIGURE 6 BEGINS
% ------------------------------------------------------------
% ------------------------------------------------------------

% ------------------------------------------------------------
% FIGURE 7 BEGINS
% ------------------------------------------------------------
% ------------------------------------------------------------

%-----------------------------------------------------------
% Section - Finite bias, temperature and magnetic field
%-----------------------------------------------------------

\section{Temperature, magnetic field, and voltage bias effects.}
\label{finiteB}

In this section we discuss how the zero-bias and zero-temperature
results are modified by considering the more general case
of a finite bias, temperature, and 
magnetic field applied perpendicular to the graphene plane. 

\begin{figure}[ht]
\begin{center}
\includegraphics*[width=.99\columnwidth]{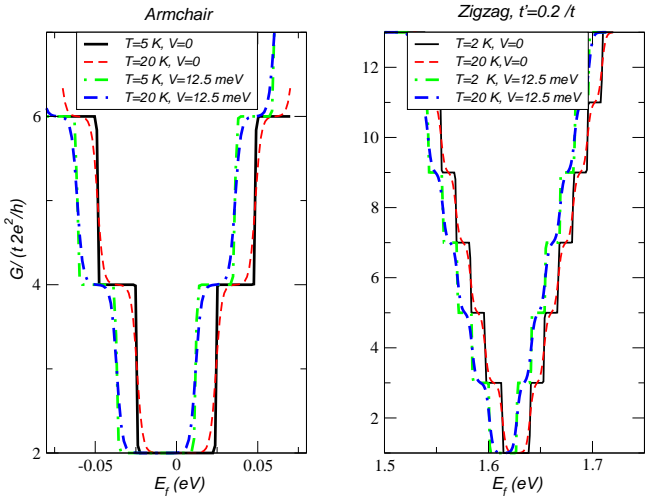}
\end{center}
\caption{(color online)
Conductance (in units of $2 \tilde t e^2/h$) for armchair and zig-zag systems as
function of $E_F$, for various values of $T$ and $V$
($t=2.7$ eV).}
\label{TB_GVT}
\end{figure}

Close to equilibrium (where $\mu_1\simeq \mu_2=E_F$)
the conductance can be obtained from  (\ref{I}) as: 
\begin{equation}
G(V\simeq 0,T)=\frac {2e^2}{h}\int d\epsilon M(\epsilon)\tilde
t(E,V) \left(- \frac{df (\epsilon-E_F)}{d \epsilon} \right)\, .
\end{equation}
For finite bias the conductance is determined from Eq. (\ref{I}) 
after a simple numerical derivative in relative to the bias
potential. 
Notice that in equilibrium, the
changes in $E_F$ can be obtained by simple changes in the
value of the gate voltage. The results
for $G(V,T)$ as function of $E_F$, $V$, and $T$, are
shown in Fig.~\ref{TB_GVT}. One can clearly sees that, as 
predicted, the conductance is quantized in units of 
$2 \tilde t e^2/h$, being even in the case of the armchair
edge and odd in the case of the zig-zag edge. The temperature 
makes the plateaus in the conductance smooth. 
Application of a bias voltage $V$ shifts the position of the 
conductance plateaus, as expected.

% ------------------------------------------------------------
% FIGURE 8 BEGINS
% ------------------------------------------------------------
% ------------------------------------------------------------

The effect of an external  magnetic field 
in the conductance quantization
of the 2D electron gas was experimentally studied in the past \cite{beenakker}
and discussed in general terms by B\"uttiker\cite{buttiker}.
 We consider only the case of systems
with zig-zag edges for simplicity since, with periodic 
boundary conditions along the $n$ direction, a unit 
cell with only two atoms can be chosen. 
In the presence of an applied perpendicular
magnetic field $B$ the hopping integrals change to: 
\begin{equation}
t_{ij}\rightarrow t_{ij}e^{i\theta_{ij}}\,,
\end{equation} 
where the phases $\theta_{ij}=2\pi\int_i^j \bm A\cdot d\bm{l}/\phi_0$,
and $\phi_0=h/e$ is the quantum of flux. The sum of $\theta_{ij}$
over a close path defined by the hopping integrals must  equal
the value $2\pi\phi_P/\phi_0=2\pi BA_P/\phi_0$, 
where $A_P$ is the area enclosed 
by the path $P$. 
For the honeycomb lattice, the equation for the amplitudes of
the tight-binding Hamiltonian (\ref{Htb}) can be written as:
\begin{eqnarray}
\epsilon a_{n,m} &=& -t [b_{n,m}+b_{n-1,m}+e^{i2\pi\phi m}b_{n,m-1}]
\nonumber\\
&-&t'[e^{i\pi\phi/3 }a_{n,m-1}+
e^{-i2\pi\phi (m-5/6)}a_{n+1,m-1}\nonumber\\
&+&
e^{-i2\pi\phi (m-1/6)}a_{n+1,m}+
e^{-i\pi\phi /3}a_{n,m+1}
\nonumber\\
&+&
e^{i2\pi\phi (m+1-5/6)}a_{n-1,m+1}+
e^{i2\pi\phi (m-1/6)}a_{n-1,m}]\,,\nonumber\\
\\
\epsilon b_{n,m} &=& -t [a_{n,m}+a_{n,m+1}+e^{-i2\pi\phi m}a_{n+1,m}]
\nonumber\\
&-&t'[e^{-i\pi\phi /3}b_{n,m-1}+e^{-i2\pi\phi (m-1/6)}b_{n+1,m-1}\nonumber\\
&+&e^{-i2\pi\phi (m-5/6)}b_{n+1,m}
+e^{i\pi\phi/3}b_{n,m+1}\nonumber\\
&+&
b_{n-1,m+1}e^{i2\pi\phi (m+1-1/6)}+
e^{i2\pi\phi (m-5/6)}b_{n-1,m}]\,,\nonumber\\
\label{tblandau}
\end{eqnarray}
where $\phi=BA_c/\phi_0$,  and $A_c$ is the area of an hexagon
($A_c=3\sqrt 3a^2/2$ with $a \approx1.4$ \AA \, in graphene). For $\phi=0$ we
obtain the results of Sec.~\ref{landauer}.

In the presence of a magnetic field the states of the bulk graphene
are described in terms of Landau levels. At low energies, when
the Dirac fermion description (\ref{dirac}) is valid, the energy
levels are given by:
\begin{eqnarray}
E_{\pm}(n)= - 3 t' + \frac{2 \alpha}{\ell^2_B} \, n \pm 
\sqrt{ \frac{\alpha^2}{\ell_B^{4}} + \frac{2 \gamma^2}{\ell_B^{2}} n}\,,
\label{ELL}
\end{eqnarray}
where we have assumed $t' \ll t$, and defined 
$\ell_B =  \sqrt{\hbar/e B}$ as the magnetic length,
$\alpha = 9t'a^2/4$, and $\gamma=3ta/2$ ($n=1,2,\ldots$ ).  
For $t'=0$, the energy levels are given by: 
$E_{\pm}(n)= \pm \sqrt 2\gamma \ell_B^{-1}\sqrt{n}$. This result 
shows that,
for the case of Dirac fermions, and unlike the ordinary 2D 
electron gas,  the Landau levels are not equally spaced
\cite{Lippmann49}. 
Notice that the cyclotron energy, $\hbar \omega_c = \sqrt{2} v_{\rm F} \hbar/l_{\rm B}$, is 
much larger than the Zeeman energy, $g \mu_B B$ ($g \approx 2$,  $\mu_B$ 
is the Bohr magneton - for $B = 12$ T, $\hbar \omega_c \approx 0.142$
 eV and $g \mu_B B \approx 7 \times 10^{-4}$ eV). Thus, we disregard the Zeeman energy in what follows. 
 
\begin{figure}[th]
\begin{center}
\includegraphics*[width=.99\columnwidth]{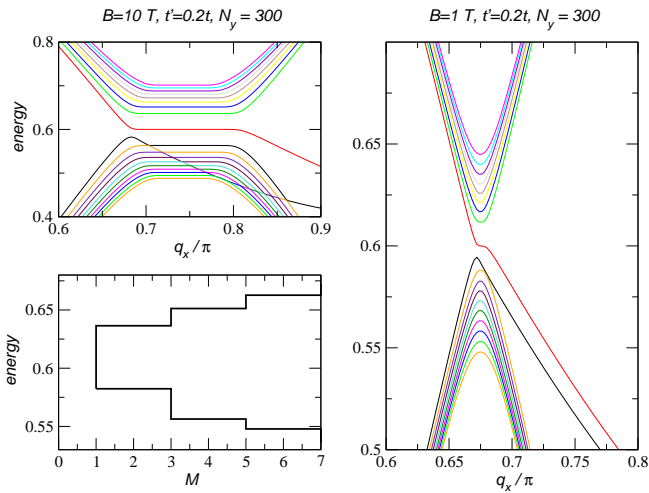}
\end{center}
\caption{(color online)
Energy spectrum (energy in units of $t$) of a 
graphene strip with a zig-zag edge
in a magnetic field $B$ (Left: $B=10$ T; Right: $B=1$ T).
Lower left panel: number 1D transverse modes $M$ for $B=10$ T.}
\label{TB_LL}
\end{figure}

In a finite system the energy levels given by 
(\ref{ELL}) are modified by the lattice structure
and by the presence of edges. This can be clearly
seen in Fig.~\ref{TB_LL}, where we plot the solution of 
(\ref{tblandau}) for a graphene strip with a zig-zag edge. 
Clearly at $B=10$ T
the Landau level structure predicted by (\ref{ELL}) is seen 
to develop close
to the Dirac point. The effect of the magnetic field on $G$
is two fold: (1) the magnetic field leads to non-dispersive 
magnetic levels, with a large degeneracy; (2) the energy 
level spacing is modified 
giving rise to a piling up of energy levels  as one moves away from the Dirac point.

% ------------------------------------------------------------
% FIGURE 9 BEGINS
% ------------------------------------------------------------
% ------------------------------------------------------------

Notice that our discussion is valid for weak magnetic
fields and hence does not apply to the quantum Hall regime
that was discussed in refs. [\onlinecite{pga,sg}]. We
observe that states with values of  $q_x$ away from the system edges can not
contribute to the conductance, since their group velocity is zero.
Only those levels having non-zero group velocity can act as 1D
channels for electron transport. The piling up of energy
levels has the experimental consequence that the observation 
of many quantized plateaus
becomes difficult. In addition, the increase of the degeneracy of each 
Landau level with the increase of the field
reduces the number of observable plateaus, as in the
normal electron gas \cite{beenakker}. On the one hand, when we compare
$M(\epsilon)$ for $B=0$ and  
$B=10$ T we see that energy width of the $M=1$ step in 
the latter case  has a much larger value. On the other hand, the piling up
of the Landau levels leads to the reduction the energy
value of plateaus
(see Figs. \ref{TB_M2} and \ref{TB_LL}). A small field $B$
does not lead  the formation of Landau levels, but removes
the degeneracy of the band formed from the flat band of zero modes
when $t'$ is considered, and leads to a $M=1$ energy step width
of a larger value when compared with the $B=0$ case.
%-----------------------------------------------------------
% Section - conclusion
%-----------------------------------------------------------
\section{summary and concluding remarks}
\label{sum}

We have discussed the tunneling transport in
clean mesoscopic graphene strips.
We show that different graphene strips have
different conductance values due to different types
of edges. As a general consequence of the graphene band structure,
and at odds to the usual 2D electron gas systems,
the conductance always increase as we move away from the
Dirac point, and therefore the conductance assumes a ``V''-shape
form as a function of the gate potential. The 
lowest value of $G/\tilde t$ in the zig-zag edge system 
is $2e^2/h$ (for a non-zero, albeit small, $t'$), 
whereas in an armchair edge system we find $4e^2/h$. 

We have studied in detail the plateaus in the conductance 
of graphene strips as a function of temperature, applied 
gate voltage, and external magnetic field. We have seen 
that the temperature smoothes out the plateaus and that 
applied gate voltages shift the plateaus in energy. 
We have also discussed the effect of next nearest neighbor 
hopping energy $t'$, that breaks the particle-hole symmetry 
of the problem and introduces dispersion for in the zero modes. 
The effect of a finite magnetic field is quite interesting in 
these systems because of the unusual relation between the 
energy and the Landau level index. We show that a magnetic 
field has effect in piling up the conductance steps and modify 
their size in energy. These effects should be easily observable 
in ultra clean mesoscopic graphene strips.

For graphene samples of 10-100 $\mu$m size \cite{pnas} it was found
that the conductivity, given by $\sigma=G L/W$, 
 has the universal value of $\sim 4e^2/h$. This result can be understood
using a bulk calculation of the effect of vacancies on the
electric linear response (Kubo formula) of Dirac fermions \cite{pga}. 
>From the point of view of coherent tunneling, these
experimental results indicate that these samples
are in the ohmic regime, having a 
mean free path shorter than the system size.  We believe,
however, that in ultra clean graphene samples it will be
possible to observe conductance quantization and interference 
patterns, as it is the case of carbon nanotubes. We hope that
our results will stimulate further studies of transport in these
amazing systems.

%-----------------------------------------------------------
% Section - Acknowledgments
%-----------------------------------------------------------
\begin{acknowledgments}
We thank A. Geim, P. Kim, and W. de Heer for stimulating discussions.
N.M.R.P. thanks ESF Science Programme INSTANS 2005-2010
and 
FCT under the grant POCTI/FIS/58133/2004.
A.H.C.N. was supported through NSF grant DMR-0343790.
\end{acknowledgments}

%************************************************************
% End of the Manuscript
%************************************************************

\end{document}